\documentclass[prd,twocolumn,nofootinbib, preprintnumbers, superscriptaddress]{revtex4-2}

\usepackage{amsmath,amssymb,slashed,braket}
\usepackage{graphicx}
\usepackage{epstopdf}
\usepackage{float}
\usepackage{amsfonts}
\usepackage[colorlinks=true,
            linkcolor=blue,
            urlcolor=blue,
            citecolor=green,          
            bookmarks=true,
            bookmarksnumbered=true,
            breaklinks=true,
            pdfpagemode=Fullscreen,
            pdfstartview=FitBH]{hyperref}

\usepackage[normalem]{ulem}
\usepackage{color}
\usepackage{aas_macros}

\definecolor{Orange}{cmyk}{0,0.61,0.87,0}
\definecolor{JungleGreen}{cmyk}{0.99,0,0.52,0}
\definecolor{OliveGreen}{cmyk}{0.64,0,0.95,0.40}
\definecolor{Brown}{cmyk}{0,0.81,1,0.60}
\definecolor{RoyalBlue}{cmyk}{0.71,0.53,0,0.12}
\definecolor{Gray}{cmyk}{0,0,0,0.40}
\definecolor{LightPink}{cmyk}{0.0,0.25,0,0}
\definecolor{LLightPink}{cmyk}{0.0,0.10,0,0}
\definecolor{LightBlue}{cmyk}{0.25,0,0,0}
\definecolor{LightGray}{cmyk}{0,0,0,0.2}

\newcommand{\f}{\mathfrak{f}}

\usepackage{xcolor}
\definecolor{gesfpurple}{rgb}{0.47,0.19,0.42}

\definecolor{gesflanse}{rgb}{0.00,0.50,0.50}

\definecolor{gesfblue}{rgb}{0.08,0.42,0.76}

\definecolor{gesfred}{rgb}{1,0,0}

\definecolor{gesfwhite}{rgb}{1,1,1}

\definecolor{gesfblack}{rgb}{0,0,0}

\newcommand{\geqn}[1]{Eq.\,\hypersetup{linkcolor=blue}(\ref{#1})\hypersetup{linkcolor=blue}}
\newcommand{\gfig}[1]{{\hypersetup{linkcolor=violet}Fig.\,\ref{#1}\hypersetup{linkcolor=blue}}}

\graphicspath{{figs/}}

\begin{document}

\title{F\'eeton ($B-L$ Gauge Boson) Dark 
Matter for the 511-keV Gamma-Ray Excess and the Prediction of Low-energy Neutrino Flux }

\author{Yu Cheng}
\email{chengyu@sjtu.edu.cn}
\affiliation{Tsung-Dao Lee Institute \& School of Physics and Astronomy, Shanghai Jiao Tong University, China}
\affiliation{Key Laboratory for Particle Astrophysics and Cosmology (MOE) \& Shanghai Key Laboratory for Particle Physics and Cosmology, Shanghai Jiao Tong University, Shanghai 200240, China}

\author{Weikang Lin}
\email{weikanglin@ynu.edu.cn}
\affiliation{South-Western Institute For Astronomy Research, Yunnan University, Kunming 650500, Yunnan, P. R. China}
\affiliation{Tsung-Dao Lee Institute \& School of Physics and Astronomy, Shanghai Jiao Tong University, China}
\affiliation{Key Laboratory for Particle Astrophysics and Cosmology (MOE) \& Shanghai Key Laboratory for Particle Physics and Cosmology, Shanghai Jiao Tong University, Shanghai 200240, China}

\author{Jie Sheng}
\email{shengjie04@sjtu.edu.cn}
\affiliation{Tsung-Dao Lee Institute \& School of Physics and Astronomy, Shanghai Jiao Tong University, China}
\affiliation{Key Laboratory for Particle Astrophysics and Cosmology (MOE) \& Shanghai Key Laboratory for Particle Physics and Cosmology, Shanghai Jiao Tong University, Shanghai 200240, China}

\author{Tsutomu T. Yanagida}
\email{tsutomu.yanagida@sjtu.edu.cn}
\affiliation{Tsung-Dao Lee Institute \& School of Physics and Astronomy, Shanghai Jiao Tong University, China}
\affiliation{Key Laboratory for Particle Astrophysics and Cosmology (MOE) \& Shanghai Key Laboratory for Particle Physics and Cosmology, Shanghai Jiao Tong University, Shanghai 200240, China}
\affiliation{
Kavli IPMU (WPI), UTIAS, University of Tokyo, Kashiwa, 277-8583, Japan}

\begin{abstract}
The f\'eeton is the gauge boson of the $U(1)_{B-L}$ gauge theory. If the gauge coupling constant is extremely small, it becomes a candidate for dark matter. We show that its decay to a pair of electron and positron explains the observed Galactic 511-keV gamma-ray excess in a consistent manner. This f\'eeton dark matter decays mainly into pairs of neutrino and anti-neutrino. Future low-energy experiments with improved directional capability make it possible to capture those neutrino signals. The seesaw-motivated parameter space predicts a relatively short f\'eeton lifetime comparable to the current cosmological constraint.
%We discuss possible detection of the f\'eeton neutrinos in low-energy neutrino detectors such as Borexino and Juno.

\end{abstract}

\maketitle 

\section{Introduction} 
The heavy Majorana right-handed neutrinos are very attractive since they naturally induce tiny neutrino masses by the seesaw mechanism \cite{Minkowski:1977sc,Yanagida:1979as, *Yanagida:1979gs,GellMann:1980vs,Wilczeck:1979CP}, and at the same time their decays in the early universe generate the Universe's baryon asymmetry through the leptogenesis \cite{Fukugita:1986hr}. It is well known that all gauge anomalies involving the $U(1)_{B-L}$ symmetry \cite{Wilczeck:1979CP} are canceled out if we introduce three right-handed neutrinos in the standard model (SM). Therefore, it is very interesting to consider the gauge $U(1)_{B-L}$ extension of the SM as the next step beyond the SM.

It has been pointed out that the $B-L$ gauge boson can be identified with the dark matter (DM) if the $B-L$ gauge coupling constant $g_{B-L}$ is sufficiently small \cite{Choi:2020kch,Okada:2020evk,Lin:2022xbu}. We call it the ``f\'eeton" or ``f\'eeton DM" \cite{Lin:2022xbu}.

%Two of the present authors (W.L and T.T.Y) has, recently, shown \cite{Lin:2022mqe} that the 511 keV gamma ray excess from the center of our galaxy can be explained by decay of the f\'eeton to a pair of electron and positron. However, there are two problems. One is the tension between the prediction of the the gamma ray flux ratio between from bulge and from disk and the observation. This tension might be solved by more details analyses of motions of the produced positrons from the f\'eeton decay. The other is that the f\'eeton density must be much lower than the DM density, otherwise the flux of 511 keV excess becomes too large and exceeds the observed flux by many order magnitude. We need invoke some other DM candidate, which weakens our motivation for the f\'eeton DM hypothesis.

Two of the present authors (W.L. and T.T.Y.) have recently shown \cite{Lin:2022mqe} that the 511 keV gamma-ray excess from the center of our galaxy \cite{1972ApJ...172L...1J,1975ApJ...201..593H,1978ApJ...225L..11L} can be explained by decay of the f\'eeton to a pair of electron and positron. It successfully explains the upper bound of the positron injection energy that is inferred from the non-detection of the 1-3 MeV 
diffused gamma ray\cite{Beacom:2005qv}. However, there are several caveats in such a scenario. First, in order to avoid an overproduction of electron-positron pairs, f\'eeton can only constitute a small fraction of DM. Second, being a small fraction of DM means that it is much more difficult to search for the signal of neutrinos decayed from f\'eeton DM. Third, it suffers from a mild tension that the resultant positron-annihilation flux is insufficiently sharp towards the Galactic center (GC), which is a common issue for all proposals with DM decays \cite{Ascaisbar:20006apil}.\footnote{However, the tension in the morphology of the signal between the observation and the prediction of DM decays is not decisive. There exist uncertainties in the transportation of positrons in the interstellar medium and more complete surveys in the disk area are still needed.}

In this paper, we consider a new scenario for forming the positronium that allows f\'eeton to be the dominant DM and significantly enhances the detectability. %, and possibly alleviates the morphology issue. 
This new scenario predicts the neutrino flux of its energy peak at 511 keV with no higher-energy continuum. We discuss how to test this neutrino with low-energy neutrino experiments like Borexino and Juno. %However, we find it difficult to discover the neutrinos produced by the f\'eeton DM decay without a very precise determination of the neutrino directions.\\

%{\bf Lin-Yanagida model for the 511 keV gamma ray excess and its problems}

\section{A new scenario for the 511 keV gamma-ray excess and consistent parameters in the f\'eeton DM model}\label{sec:new_scenario}
In this section, we discuss a new parameter region of the original f\'eeton DM model \cite{Lin:2022xbu}. The low-energy physics is described by only two free parameters, the mass $m_\f$ and the gauge coupling constant $g_{B-L}$ of the f\'eeton. 
These two parameters are related by $m_\f = 2g_{B-L} V_{B-L}$, where $V_{B-L}$  is the Vacuum Expectation Value (VEV)) of a Higgs boson $\Phi$ with a $B-L$ charge of $+2$.
Here, the right-handed neutrinos, $N_i~~ (i=1-3)$, acquire Majorana masses of $h_i V_{B-L}$ with the constant parameters $h_i$ defined by the Yukawa interactions $\frac{1}{2}h_i \Phi N_iN_i$. We have assumed all leptons including the right-handed neutrinos have the $B-L$ charge of $-1$ \footnote{The definition of $B-L$ charge has an ambiguity related to the $U(1)$ hypercharge gauge transformation \cite{Okada:2020evk}. However, our main conclusions are not much changed except for a very special parameter region where the f\'eeton coupling to the electron and positron pair is suppressed  \cite{yanagida:2024FM}. }.

The decay rate $\Gamma_{\mathfrak{f} }$ of the f\'eeton is given by
\begin{equation}
\Gamma_{\mathfrak{f} }
= \frac{g_{B-L}^2 m_{\mathfrak{f}}}{24 \pi}
    \left[ 
    3 +
    2 \sqrt{1-\frac{4 m_e^2}{m_{\mathfrak{f}}^2}}
    \left(1+\frac{2 m_e^2}{m_{\mathfrak{f}}^2}\right)\right],
\end{equation}
where we assume 
$m_{\mathfrak{f}}>2m_e$ so that both the decays to the neutrino-anti-neutrino pairs and the electron-positron pair are included, 
$m_e$ is the electron mass and the neutrino masses have been ignored.  %This positron is supposed to form the positronium producing the observed 511 keV gamma ray. 
%We assume that the f\'eeton is the dominant DM in the present universe. The production mechanism of the f\'eeton DM will be discussed in the last section of this letter. 
If the f\'eeton is the dominant DM with 
$m_\mathfrak{f} > 2m_e + 13.6\,$eV, the positrons produced from f\'eeton decay can 
form positronium via the
charge exchange of positrons with hydrogen atoms. 
Such positroniums eventually annihilate into 
gamma rays.
However, the electron-positron pairs 
would be over-produced and 
the resultant Galactic 
511 keV gamma ray would 
largely exceed the  
observation \cite{Lin:2022mqe}. 
%\sout{Then, we see the reason why the f\'eeton cannot be the dominant DM. The electron-positron pairs would be over-produced and the resultant Galactic 511 keV gamma ray would largely exceed the  observation as long as the branching ratio of the $\mathfrak f \rightarrow e+{\bar e}$ decay channel is $O(1)$ as pointed out in \cite{Lin:2022mqe}.}

Now we are at the main point of this paper. If the f\'eeton mass $m_{\mathfrak{f}}$ is very close to the threshold of the decay to the electron-positron pair, that is, $m_{\mathfrak{f}}\simeq 2 m_e$, such a decay is strongly suppressed and the branching ratio to the $e+{\bar e}$ final state becomes very small so that the predicted excess of the 511 keV gamma ray can be consistent with the observation while the f\'eeton is the dominant DM. However, it does not look successful. It is usually assumed that the intermediate state of positronium is formed by the produced positrons by charge exchanges with neutral hydrogen atoms \cite{Beacom:2005qv}. This can happen only if the kinetic energy of the positron is larger than the corresponding threshold 6.8\,eV---the difference in the binding energy between neutral hydrogen and positronium. This sets the lower bound of the f\'eeton mass and hence the $e+{\bar e}$ branching ratio.

However, if the produced positrons have a kinetic energy smaller than 6.8\,eV, instead of with neutral hydrogen atoms, they can predominantly form positronium with free electrons in warm or hot ionized gas; see, e.g., Fig.\,27 in \cite{Prantzos:2010wi}. 
%Such an environment can be realized close to the Milky Way bulge.\footnote{The fraction of the ionization in the Disk is about 20-40 $\%$ \cite{}. Thus, it may be not easy to explain the observed morphology of the 511 keV signal.} 
This possibility has been largely overlooked in the literature. Thus, we consider the following scenario. The f\'eeton constitutes the total DM and has a mass very close to the threshold of the decay to an electron-positron pair, more explicitly, $2m_e<m_{\mathfrak{f}}<2m_e+13.6$\,eV. They decay to electrons and positrons with a small branching ratio compared to the decay to neutrinos and anti-neutrinos. The produced positrons mainly form positroniums with free electrons in ionized environments, which subsequently annihilate to produce the 511 keV gamma rays. %In this way, the tension in the morphology of the 511 keV gamma-ray signal between the observation and the prediction of the f\'eeton decays is also solved as we pointed in the introduction. 
%As we shall explain, such a scenario overcomes several difficulties in previous studies and predicts a neutrino signal that is testable in the near future with low-energy neutrino experiments such as Borexino and JUNO.

%%%%%%%%%%%%%
Assuming there are sufficient free electrons so that the positron produced from f\'eeton decays annihilate immediately and the gamma-ray emission rate traces the f\'eeton decay rate. The angular differential gamma-ray flux is given in \cite{Lin:2022mqe}. In the first order of $\Delta m/m_e$, it reads,
\begin{align}
    \frac{{\rm d}\Phi_{511}}{{\rm d}\Omega}
    &=4\times10^3\left(\frac{g_{B-L}}{10^{-20}}\right)^2\left(\frac{\Delta m}{m_e}\right)^{1/2} \nonumber\\
    &~~~~~~~~~~~~\times  \tilde{D}_{\rm N}(\cos\theta)\,[{\rm cm}^{-2}{\rm s}^{-1}{\rm sr}^{-1}]\,,\label{eq:positron-annihilation}
\end{align}
where $\Delta m\equiv m_{\mathfrak{f}}-2m_e$, $\tilde{D}_{\rm N}(\cos\theta)$ is a function that represents the morphology of the flux and is normalized so that $\smallint \tilde{D}_{\rm N}\,{\rm d}\Omega=4\pi$.
The angle between the direction of the flux and GC $\theta$
is defined by 
$\cos\theta \equiv \cos b\cos\ell$ with galactic coordinates $(b, \ell)$. 
The morphology function $\tilde{D}_{\rm N}(\cos\theta)$ depends on the 
Galactic DM distribution 
which we have adopted an NFW 
profile \cite{Navarro:1995iw}. 

\begin{figure}
    \centering
    \includegraphics[width=0.486
    \textwidth]{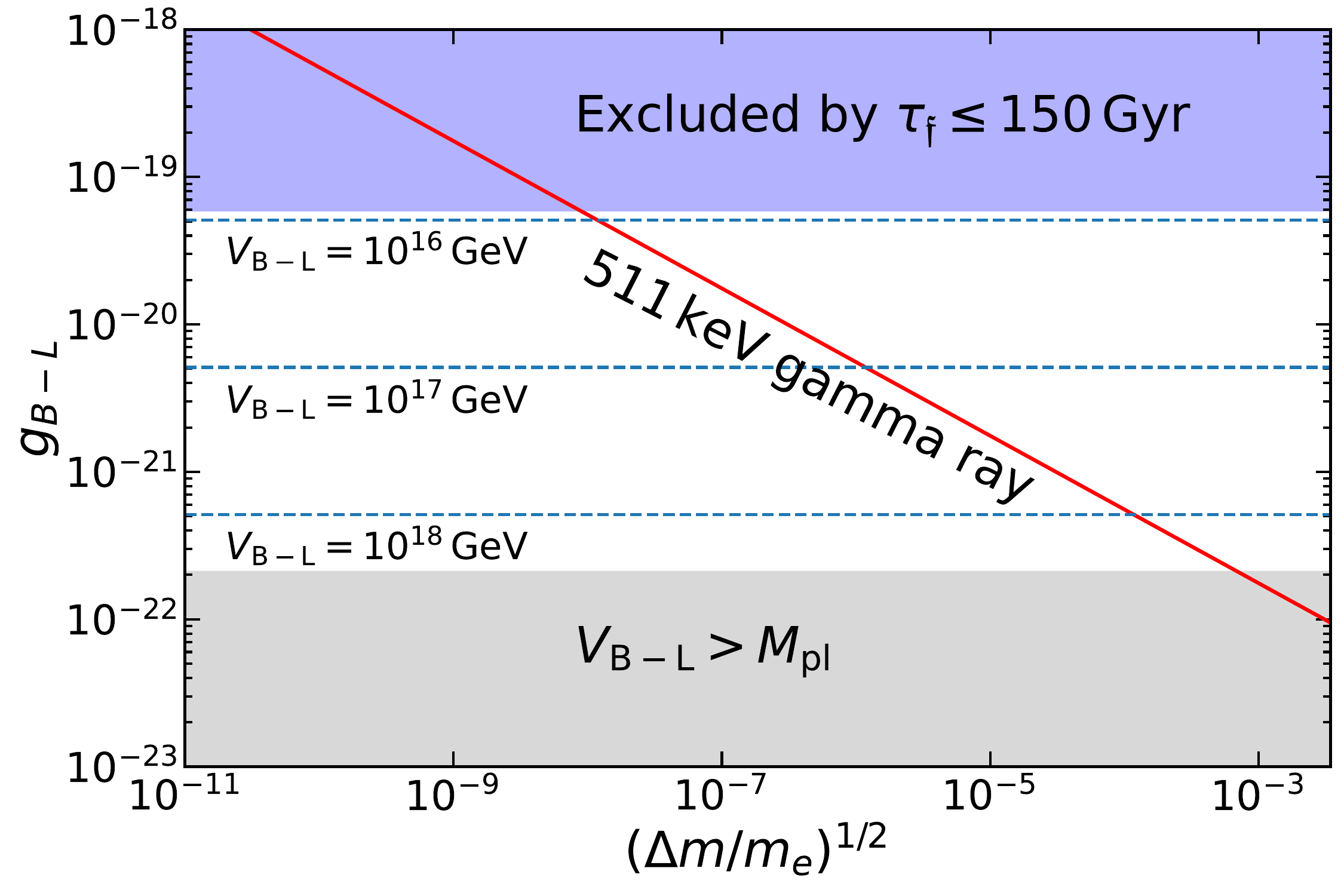}
    \caption{The red line shows the corresponding $g_{B-L}$ to explain the observed 511 keV gamma ray excess. 
    The blue-shaded region is excluded 
    by supposing the lifetime of 
    f\'eeton should be larger than 
    ten times of the universe age, $\tau_{\mathfrak{f}} > 150\,$Gyr. 
    The gray-shaded region 
    represents for $V_{B-
    L}>M_{\rm pl} = 2.4 \times 
    10^{18}$\,GeV.
     }
    \label{fig:511Constraint}
\end{figure}

Recall that  Ref. \cite{Siegert:2015knp} measures a 511-keV gamma-ray intensity of $0.96\pm0.07$\,cm$^{-2}$\,s$^{-1}$ 
from the bulge region with a Full-Width-at-Half-Magnitude (FWHM) of $20.55$ degrees. 
To proceed, we integrate \eqref{eq:positron-annihilation} within $\theta<10.28$ degrees from GC and set it as half of the measured bulge flux to be the observed bulge 511-keV gamma-ray intensity. This gives us the f\'eeton DM parameter space that can explain the observed 511-keV gamma ray, which is shown as the red line in Fig.~\ref{fig:511Constraint}. We also show the parameter space excluded by the lifetime constraint (purple) and that with a super-Planckian VEV (gray). As expected, the f\'eeton mass should be close to twice the electron mass. The closer $m_\f$ to $2m_e$, the larger the gauge coupling constant should be to account for the 511 keV gamma ray intensity. For the maximally allowed gauge coupling constant, i.e., $g_{B-L}\simeq6\times10^{-20}$, we require $(\Delta m/m_e)^{1/2}\simeq10^{-8}$. The mass difference $\Delta m$ is identified to be the total kinetic energy of the electron-positron pair produced by the f\'eeton decay and thus the positrons are non-relativistic \footnote{An alternative way to suppress the f\'eeton decay to the electron and positron is given by the mechanism mentioned in footnote 2. In this case, it might be interesting that the anti-neutrinos produced from the f\'eeton decay causes the 511 keV gamma ray emissions in the detector. However, it will be not easy to distinguish the signal from the geo neutrinos.}.

The above has demonstrated that the f\'eeton can constitute the total DM while being able to explain the intensity of the 511-keV gamma ray considering that the non-relativistic positrons produced from f\'eeton decays form positronium with free electrons in ionized environments. Since now the f\'eeton is not merely a tiny fraction of DM, it significantly enhances the neutrino and anti-neutrino fluxes produced from f\'eeton decays as compared to the scenario in the previous work \cite{Lin:2022mqe}. In addition, the neutrino and anti-neutrino energies peak at 511 keV, which is above the threshold for directional determination by current detectors that use Cherenkov lights. Thus, the new scenario in this work has much better detectability compared to that in \cite{Lin:2022xbu}.

%We see that the gauge coupling constant $g_{B-L}$ becomes larger if we take the mass difference $m_\f-2 m_e$ smaller. Notice that the neutrino flux becomes larger when the B-L gauge coupling constant $g_{B-L}$ becomes larger. \\
%In the next section we take a reasonable value of the mass difference as an example and discuss how to detect the neutrinos produced by the f\'eeton decay, in future low-energy neutrino experiments such as Borexino and Juno.\\

\section{Possible detection of the predicted neutrinos at the low-energy neutrino experiments}

\subsection{Low-energy neutrino detectors}\label{sec:low-energy-detectors}
The main decay mode of the f\'eeton is that to the neutrino-anti-neutrino pair. The energy of the resultant neutrinos is peaked at $\simeq 511$ keV because the f\'eeton mass is just above the threshold of the decay, that is, $m_{\mathfrak f} \simeq 2 m_e$ as explained in the previous section. The neutrino flux from the Galactic and the extra-galactic f\'eeton DM decay have been calculated in \cite{Lin:2022xbu}. In this section, we discuss the detectability of the corresponding neutrino flux by setting the f\'eeton mass to $\simeq1$ MeV and taking the optimal coupling constant allowed by the DM lifetime constraint, i.e., $g_{B-L}=5.85\times10^{-20}$; see Ref.,\cite{Lin:2022xbu}. This value of $g_{B-L}$ is also the most motivated value as we shall explain.
%of the 511 keV neutrino from the f\'eeton decay in the Milkey Way.% which is enough for discussing detection of the f\'eeton neutrinos.

%The flux of the neutrinos are predicted by the $B-L$ gauge coupling constant, $g_{B-L}$. 
%he solar neutrino detection experiments such as the Borexino can have a possibility to detect the neutrinos produced by the  f\'eeton decay. 
We first discuss the detectability for solar neutrino experiments since the predicted energy falls into their target energy range. Adopting Eq.\,(5) in \cite{Lin:2022xbu} with $g_{B-L}=5.85\times10^{-20}$, we see that 
the neutrino flux of the f\'eeton decay is $\simeq3.4\times10^6$\,cm$^{-2}$s$^{-1}$, which is three-order-of-magnitude smaller than the $^7$Be neutrino flux \cite{Bahcall:2004fg} although the peak energy is different. %Therefore, it seems very difficult to detect it in the near future solar neutrino experiments. 
We then evaluate the electron recoil spectrum per unit mass of liquid scintillator as the neutrinos from the f\'eeton decay scatter off electrons in neutrino detectors. The energy differential event count is, 
\begin{equation}
    \frac{dN}{dE_e}
    =
    \int d E_{\nu} 
    \frac{d \Phi_{\nu}}
    {dE_{\nu}}
    \cdot
    \frac{d \sigma}{dE_{e}}(E_\nu)
    \cdot
    N_T \cdot t.
\end{equation}
Here, $E_e$ is the kinetic energy of the recoil electron, $\frac{d \Phi_{\nu}}{dE_{\nu}}$ is the neutrino flux from the f\'eeton decay \cite{Lin:2022xbu,Lin:2022mqe}, $N_T = (3.307 \pm 0.015) \times 10^{29}$ is the number of electrons per ton of the liquid scintillator \cite{Kumaran:2021lvv}, $t$ is the exposure time which we take one year for illustration. Besides, 
the neutrino-electron scattering differential cross section is given by \cite{tHooft:1971ucy}, 
\begin{multline}
\frac{\mathrm{d} \sigma}{\mathrm{d} E_e}
=
\frac{G_F^2 m_e}{2 \pi} 
\left[ \left(g_{\mathrm{V}}+g_{\mathrm{A}}\right)^2+\left(g_{\mathrm{V}}-g_{\mathrm{A}}\right)^2\left(1-\frac{E_e}{E_\nu}\right)^2 \right.\\
\left.-\left(g_{\mathrm{V}}^2-g_{\mathrm{A}}^2\right) \frac{m_e E_e}{E_\nu^2}\right]\,.
\label{eq:NeutrinoECrossSection}
\end{multline}
For the electric neutrino-electron scattering, $g_V = 1/2 + 2 e^2/g^2$ and $g_A = 1/2$. For the $\nu_\mu$- and $\nu_\tau$-electron scatterings, $g_V = (-1/2+ 2 e^2/g^2)$, $g_A = -1/2$ where $g$ is the weak gauge 
coupling. For the anti-neutrino electron scatterings, the corresponding cross sections can be obtained by replacing $g_A$ with $-g_A$ in \geqn{eq:NeutrinoECrossSection}. 

\begin{figure}
    \centering
    \includegraphics[width=0.486
    \textwidth]{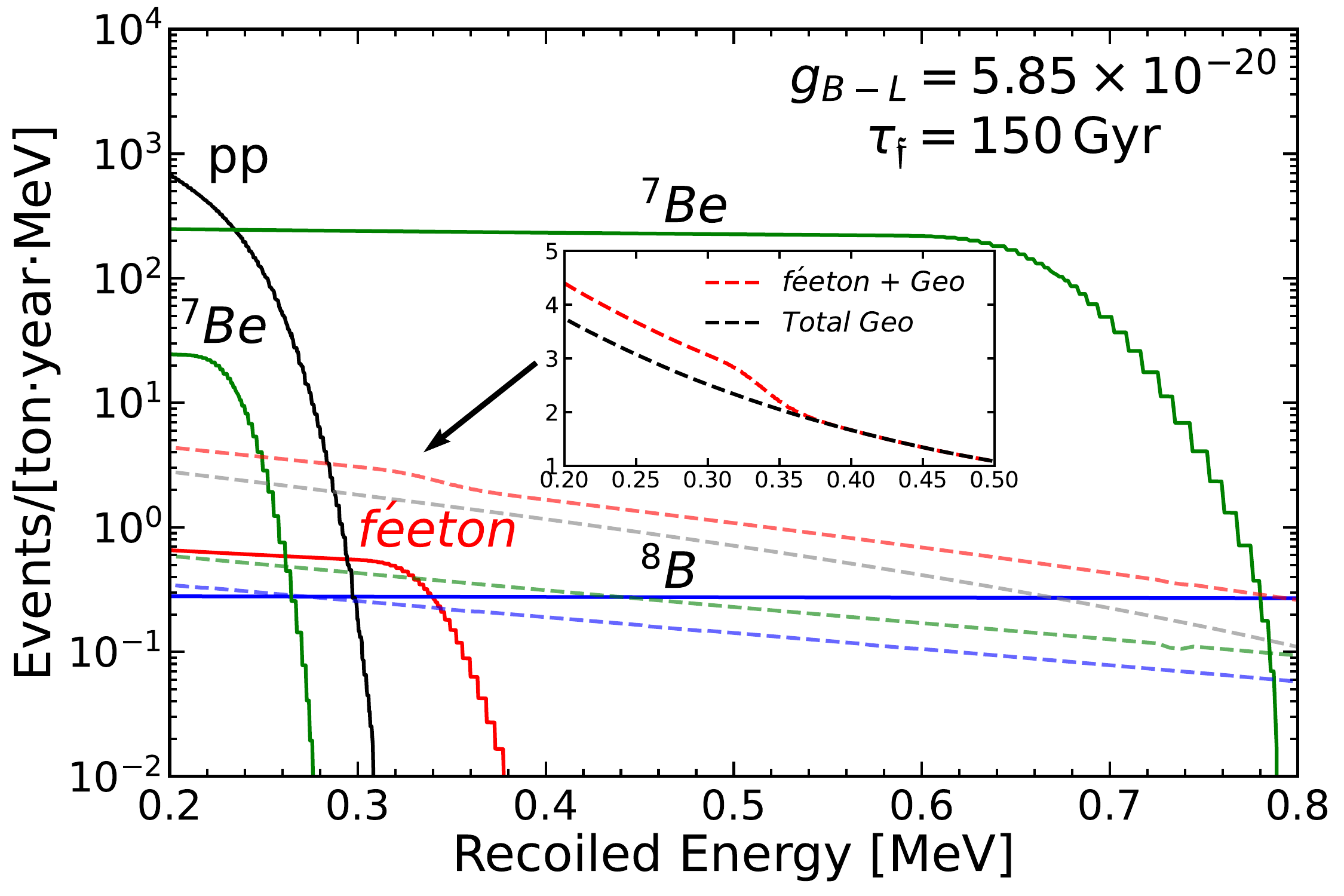}
    \caption{Electron recoil energy spectra caused by different neutrino sources.   
    The red line shows the electron recoil spectrum for maximum neutrino flux from f\'eeton decay by assuming  $\tau_{\mathfrak{f}} = 150$\,Gyr which corresponds to $g_{B-L} = 5.85 \times 10^{-20}$. 
    The electron recoil spectrum from the solar $pp$ ($^7$Be,$^8$B) 
    is shown as the black (green, blue) line. 
    The geo-neutrino contribution from K (U, Th) decay is shown as a gray (green, blue) dashed line. 
    The red-dashed curve shows the spectrum for the total geo-neutrinos plus the f\'eeton neutrinos. For comparison, the total geo-neutrinos background is shown as the black dashed line in the subfigure.
     }
\label{fig:NuElectronRecoil}
\end{figure}

The resultant electron recoil spectrum caused by the neutrino flux from the f\'eeton decay is shown by the red line in \gfig{fig:NuElectronRecoil}, along with those for the solar neutrinos of different channels.
%We take the optimal $g_{B-L} = 5.85 \times 10^{-20}$ which corresponds to $\tau_{\mathfrak{f}} = 150$\,Gyr to maximum the neutrino flux.
We can see the spectrum for the f\'eeton case is of the same order as that of $^{8}$B and almost three-order-of-magnitude smaller than $^{7}$Be. % The direction determination will need to be discussed in order to continue.

%It is very interesting that the gauge coupling constant $g_{B-L}$ is determined so that the f\'eeton DM decay can explain the observed 511 keV gamma-ray excess from the bulge in our galaxy. As shown in \gfig{fig:511Constraint}, the gauge coupling constant is fixed by the mass of the f\'eeton and hence all low-energy physics is given by the precise value of the f\'eeton mass, $m_\f$.

The most serious problem, however, is the presence of several isotope decays inside the scintillators in low-energy solar neutrino experiments such as Borexino. For example, the beta decay of $^{210}$Po produces neutrinos whose energy range 
covers 511 keV and flux is $7 \sim 8$ order magnitude larger than the f\'eeton neutrino flux \cite{Bellini:2011rx,BOREXINO:2018wno}. Removing such contamination is crucial for the search of the f\'eeton neutrino flux. In this paper we will not discuss the purification mechanism \cite{Bellini:2016kau} of such isotopes and only assume that we can remove such background neutrinos at a sufficient level for the detection of the f\'eeton neutrinos and discuss possible discrimination of f\'eeton signals from the solar neutrinos, mainly $^{7}$Be neutrinos. 

Another problem is the geo-neutrino contamination. Geo-neutrinos consist of electron-type anti-neutrinos produced by the beta decay of radionuclides, mostly ${}^{40}$K, ${}^{232}$Th, and ${}^{235}$U, in the Earth \cite{Vitagliano:2019yzm}. Their energies range from $0.1$\,MeV to $3$\,MeV. As the dashed lines shown in \gfig{fig:NuElectronRecoil}, the geo-neutrino contributions for the electron recoils are larger than that for the f\'eeton neutrino. %These fluxes do not have any annual modulation and hence it is very hard to discriminate them. 
However, it is still possible for a Borexino-type experiment to distinguish it through direction information.

\begin{figure}
    \centering
    \includegraphics[width=0.99\linewidth]{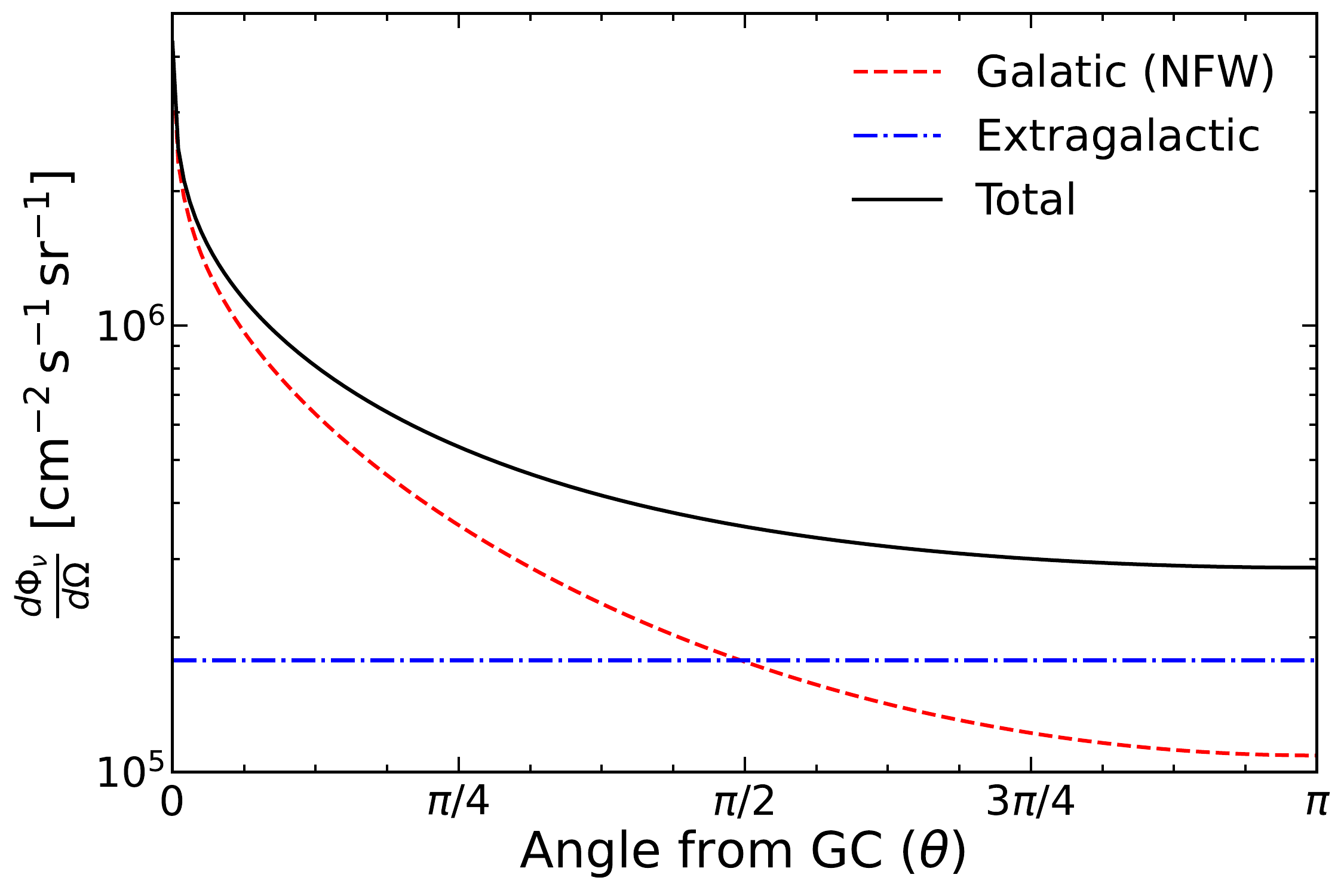}
    \caption{ The angle 
    distribution for neutrino flux 
    from f\'eeton DM decay 
    with mass 
    $m_\f \simeq 2 m_e$
    and coupling 
    $g_{\rm B-L} 
    = 5.85 \times 10^{-20}$. 
    The red dashed line shows 
    the Galactic contribution 
    with NFW profile and
    the blue dash-dotted line
    shows the extra-galactic contribution. 
    Their combination 
    is the black solid curve.
     }
     \label{fig:NFluxTotal}
\end{figure}

To mitigate the problems from the solar neutrinos and geo-neutrinos, one may utilize the direction dependence and the time modulation of neutrino fluxes. The solar neutrino flux comes from the sun and our f\'eeton neutrino flux is more concentrated towards GC. Fig.\,\ref{fig:NFluxTotal} shows the angular dependence of the f\'eeton neutrino flux assuming the f\'eeton DM distribution follows an NFW profile \cite{Navarro:1995iw} in our galaxy, where $\theta$ is the angle between the line of sight and GC. The f\'eeton neutrino flux from GC is the most intense and thus allows the highest directionality. In \gfig{fig:NFluxTotal} we also include the uniform neutrino flux decayed from the extragalatic f\'eeton DM.

Some recent solar neutrino detectors already equip direction determination capability. For instance, the Borexino experiment combines both the water Cherenkov and liquid scintillator detectors, it gathers the first directional measurement of sub-MeV solar neutrinos\cite{BOREXINO:2021xzc}. %In water Cherenkov detectors, neutrinos scatter off electrons in the medium. 
Light signals are produced via neutrinos scattering off electrons. For water Cherenkov detectors, if the recoil electron moves faster than the speed of light in water, the directional Cherenkov light is produced. Since the speed of light in water is around $v_c \simeq 0.75c$, the kinetic energy of the recoil electron should be larger than $E_e \simeq 0.5 m_e$ to produce Cherenkov lights. The corresponding minimum energy of the neutrino is $420\,$keV. Thus, the energy of the f\'eeton neutrinos is above the threshold, which allows us to make direction determination via water Cherenkov detectors. On the other hand, an electron with a kinetic energy of $0.5 m_e$ is relativistic. Such a large recoil energy can only produced by an almost-forward scattering by a 511-keV neutrino. Therefore, for the f\'eeton neutrino case, the direction of the recoil electron almost tracks the direction of the incoming neutrino.

Since the Earth is orbiting the sun, the direction of the sun is changing while the direction of GC is not. The sun and galaxy center have nearly opposite directions related to the Earth around June, which makes f\'eeton neutrino flux the most distinguishable from the solar neutrinos in terms of direction. Similarly, geo-neutrinos only come from the inside of the Earth. Based on the GC location around June, a detector should be put in the south hemisphere so that the geo-neutrinos and the f\'eeton neutrinos from the GC region are in opposite directions. 

The above discusses the difference in the direction of each signal. However, the geo-neutrinos are much more diffuse than the solar neutrinos and are more difficult to get rid of using direction determination. Fortunately, the total geo-neutrino intensity is only several times larger than that of the f\'eeton neutrinos. Suppose the solar neutrino signal can be sufficiently identified and subtracted using direction determination, we obtain the electron recoil spectrum for the total geo-neutrinos and f\'eeton neutrinos which we show by the red-dashed curve in \gfig{fig:NFluxTotal}. We can see a characteristic kink at $\sim0.35$\,MeV. This feature at that specific energy is so far unique for all known neutrino sources. While there is some uncertainty in the geo-neutrino intensity, this feature persists in the otherwise smooth geo-neutrino electron-recoil spectrum as long as the 511-keV f\'eeton neutrinos exist. Detecting such a kink will be a smoking gun to the 511-keV neutrinos and the f\'eeton DM.

%In that case, we can divide all directions into two parts, one is facing towards the sun (Part A) and another is opposite to the sun (Part B). The electron scatterred off by solar neutrino can only leave signals in Part B. However, once the sun and galaxy center has a opposite relative position to our earth and we see the electron signals in Part A, we can claim this signal is from f\'eeton without solar neutrino background.

%A similar method can be applied for the discrimination of geo-neutrinos since they come only from inside the Earth. However, those methods work when we have very high resolution of the neutrino directions and the present Borexino detector is not sufficient. The test of f\'eeton scenario requires some new experimental methods to determine the neutrino direction with a very high resolution \gred{\cite{Theia:2019non}}.

%\gblue{
%Another interesting search for the f\'eeton neutrino is given by the measurement of anti-neutrinos since the solar neutrinos don't have the anti-neutrinos. An example has been proposed for the measurement of the cosmic anti-neutrino \cite{}. It uses the capture of the electron-type anti-neutrino on the $^{163}$Ho atom as
%\begin{equation}
%    {\bar \nu_e} ~+~ ^{163}Ho \rightarrow....
%\end{equation}
%Unfortunately, this process is not applicable for the present f\'eeton anti-neutrino, since the energy 511 keV of the anti-neutrino is too large.............\\
%}
%\subsection{Low-energy anti-neutrino detectors}

\subsection{Cosmological probes}
Taking $m_\f\simeq 2m_e$, the DM lifetime constraint sets an upper bound of $g_{B-L}\lesssim6\times10^{-20}$ as explained in Ref.\,\cite{Lin:2022xbu}. On the other hand, we still have a large parameter space on the lower bound of the gauge coupling constant $g_{B-L}$, which is determined by the upper bound of the $B-L$ breaking scale. Provided a conservative upper bound $V_{B-L} <M_{Pl}$ where $M_{Pl}\simeq2.4\times10^{18}$\,GeV is the reduced Planck mass, we have $g_{B-L} \gtrsim2\times 10^{-22}$ using $m_\f=2g_{B-L}V_{B-L}$ and $m_\f\simeq2m_e$. 

However, not all of the parameter space between $6\times10^{-22}\lesssim g_{B-L}\lesssim2\times10^{-20}$ is equally motivated. It is remarkable that the observed neutrino mass $m_{\nu} \simeq 0.05\,$eV naturally predicts the $B-L$ breaking scale $V_{B-L}$ to be $O(10^{16})$\,GeV \cite{Buchmuller:1998zf} provided that all the Yukawa coupling constants for the third family right-handed neutrino $N_3$ such as $h_3$ are $O(1)$. 
%If it is the case, we predict the f\'eeton mass $\leq 1$\,MeV taking the lifetime of the f\'eeton is longer than 150Gys  (see Fig.1 in \cite{Lin:2022xbu}).
Taking that $V_{B-L}\simeq10^{16}$\,GeV and that $m_\f\simeq2m_e$ required to explain
the Galactic 511 keV gamma-ray excess, we have $g_{B-L}\simeq 5 \times10^{-20}$. Thus, the value of $g_{B-L}$ we took in Sec.\,\ref{sec:low-energy-detectors} is not merely an optimal coupling constant, but also a very motivated one. More importantly, with these values of $g_{B-L}$ and $m_\f$,
we predict that the f\'eeton DM lifetime to be $\sim150$\,Gyr. This is very encouraging because it is already close to the cosmological constraint on the DM lifetime obtained with the cosmic background and large-scale-structure (LSS) probes \cite{DeLopeAmigo:2009dc, Audren:2014bca, DES:2020mpv, Enqvist:2019tsa}. Therefore, the f\'eeton DM scenario proposed in this work implies that the cosmological effects due to the f\'eeton DM decay can soon be detected via cosmological probes with the near-future galaxy surveys such as LSST\footnote{\href{http://www.lsst.org/lsst}{http://www.lsst.org/lsst.}}, DESI\footnote{\href{https://www.desi.lbl.gov/}{https://www.desi.lbl.gov/.}}, Euclid\footnote{\href{http://sci.esa.int/euclid/}{http://sci.esa.int/euclid.}} and WFIRST\footnote{\href{https://www.skatelescope.org}{https://www.skatelescope.org.}}, and the on-going mission JWST\footnote{\href{https://webb.nasa.gov}{https://webb.nasa.gov.}}.

\section{Discussion and Conclusions}
In this paper, we have shown a scenario in which the f\'eeton is the dominant DM and consistently explains the observed magnitude of the Galactic 511-keV gamma-ray excess by the f\'eeton DM decay into electron and positron. The scenario pins down the mass of the f\'eeton very close to $2m_e$ so that the electron and positron produced by the f\'eeton decay are highly non-relativistic. The injection kinetic energy of the positrons is less than $13.6$\,eV so that they do not form positronium via charge exchanges with neutral hydrogen atoms, which is usually taken in the literature. Rather, we consider that they form positronium with free electrons in ionized environments. Different from the previous cases \cite{Lin:2022xbu,Lin:2022mqe}, the scenario in this work produces larger neutrino and anti-neutrino fluxes with an energy that is high enough for direction determination with the current solar neutrino experiments. In the future, solar neutrino experiments with improved angular resolution are promising to detect the neutrinos decayed from the f\'eeton DM. The parameter space that is consistent 
with the high-scale Seesaw mechanism also predicts 
that the f\'eeton lifetime is close to the current cosmological constraint. Thus, it implies the effects on the cosmic background evolution and LSS should be soon detected via near-future galaxy surveys.
%hence, the defused gamma ray from their motions must be highly suppressed. This will be a crucial test of the present scenario. An upcoming telescope, 

\begin{figure}[!t]
    \centering
    \includegraphics[width=0.99\linewidth]{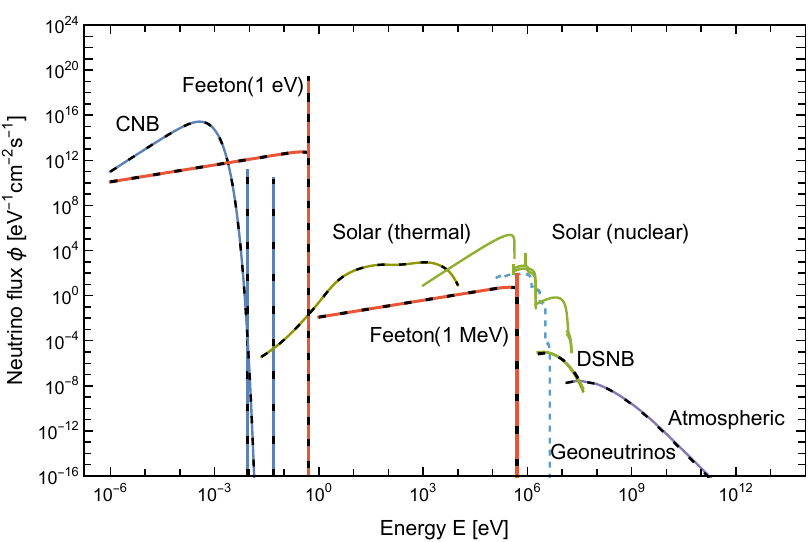}
    \caption{
    The energy spectra of neutrino flux from different sources with different colors
    \cite{Vitagliano:2019yzm}. Solid lines represent for neutrino flux while dashed lines for anti-neutrino. 
    Among them, the red lines show the neutrino flux from f\'eeton decay with taking f\'eeton mass to be $m_\f = 1\,$eV and $1\,$MeV. The galactic signal is a delta-line spectrum with total flux $3.5 \times 10^{12}\,$cm$^{-2}$s$^{-1}$ for $m_\f = 1\,$eV and $3.5 \times 10^{6}\,$cm$^{-2}$s$^{-1}$ for $m_\f = 1\,$MeV. We convolute the delta function spectrum from 1 MeV f\'eeton decay with the energy resolution of Borexino experiment at MeV region,
    $\Delta E / E = 5$\%. The 
    continuous spectrum corresponds to the extra-galactic signal.
    Note that the two blue vertical lines correspond to the two heavy generations of neutrinos in CNB.
     }
     \label{fig:AllnuFlux}
\end{figure}

Since the produced positrons are highly non-relativistic, it implies there is no higher-energy continuum in the gamma-ray excess that would be produced as high-energy positrons fly through the Galactic medium. This provides another crucial test of the scenario. Future experiments, such as the Compton Spectrometer and Image (COSI) mission that aims at detecting the soft gamma-ray of $0.2 \sim 5\,$MeV \cite{Tomsick:2021wed} will give us strong constraints on the present f\'eeton DM model.

It is assumed in this work that f\'eeton is the dominant component of the DM in the present Universe. Normally, it seems very difficult to produce the f\'eeton in the early Universe, since its gauge coupling constant $g_{B-L}$ is extremely small. 
However, it is pointed out that the light gauge bosons $\cal{V}$ can be produced abundantly during the inflation independent of the magnitude of their gauge coupling constant. 
In fact, the abundance $\Omega_{\cal V}$ is given by \cite{Graham:2015rva},
\begin{equation}
    \Omega_{\cal V}  
    \simeq 0.3 
    \left(\frac{m_{\cal V}}{6\times10^{-6}\,{\rm eV}}
    \right)^{1/2}
    \left(\frac{H_{\rm inf}}{10^{14}\,{\rm GeV}}
    \right)^2,
\end{equation}
where $m_{\cal V}$ and $H_{\rm inf}$ are the mass of the gauge boson and the Hubble constant during inflation, respectively. 
With a f\'eeton mass $m_{\mathfrak f} \simeq 1$\,MeV and setting $\Omega_{\cal V}=\Omega_\f\simeq0.25$, we predict $H_{\rm inf} \simeq 1.6\times 10^{11}$\,GeV, which is a rather low-energy inflation. Recall that $H_{\rm inf}\simeq2.6\,r^{1/2}\times10^{14}$\,GeV assuming a slow-roll inflation \cite{Planck:2018jri}, where $r$ is the primordial tensor-to-scalar ratio. The ratio $r$ is then predicted to be very low, i.e., $r\simeq4\times10^{-7}$. Thus, if the next-generation Cosmic Microwave Background (CMB) experiments detected the primordial B-mode polarization, the above f\'eeton DM production scenario would be falsified and either the f\'eeton was not the dominant DM or it was produced by other mechanisms \cite{CMB-S4:2022ght}.

Finally, we would give a general comment on the f\'eeton DM hypothesis. Figure \ref{fig:AllnuFlux} shows different neutrino fluxes from the known sources in increasing order of energy \cite{Vitagliano:2019yzm}. It contains from the cosmic neutrino background (CNB) of $10^{-6}$\,eV to the atmospheric neutrino of $10^{12}\,$eV. However, there is a blank between 0.1 eV and 1 keV.  Although different from the scenario proposed in this work, it would be interesting for the f\'eeton neutrino to be a candidate to fill the blank if the f\'eeton mass is $m_\f = 0.1$\,eV $\sim 1$\,keV, which can be achieved with a low-energy Seasaw mechanism. Taking $m_{\mathfrak f} = 1\,$eV for example, the neutrino contribution from the  f\'eeton decay is shown as the red solid curve in \gfig{fig:AllnuFlux}, while the anti-neutrino contribution with the same flux is shown as the dashed curve. %Since the DM mass is light, the large density leads to a higher neutrino flux compared with the CNB.
An interesting search for the f\'eeton neutrino is given by the measurement of anti-neutrinos.
%, since the solar neutrinos doesn't have the anti-neutrinos. 
An example has been proposed for the measurement of the cosmic anti-neutrino \cite{Vignati_2012,Gastaldo:2017edk}. It uses the capture of the electron-type anti-neutrino on the $^{163}$Ho atom,
\begin{equation}
    {\bar \nu_e} ~+~ ^{163}{\rm Ho} \rightarrow ~^{163}{\rm Dy} ~+~ E_i,
\end{equation}
where $E_i$ is some binding energy from the de-excitation of the Dy atom. 
Based on the red dashed curve in \gfig{fig:AllnuFlux}, the integrated anti-neutrino flux from f\'eeton decay with $m_{\mathfrak f} \sim 1$\,eV is larger than that of CNB. This means the capture rate of the f\'eeton anti-neutrino on $^{163}$Ho is larger than
the capture rate of the cosmic anti-neutrino. So the low-mass f\'eeton can be tested in such experiments. 
Furthermore, it would be interesting to investigate if the f\'eeton DM scenario can fit the recent observation of $511\,$keV emission from dwarf spheroidal galaxies\cite{Siegert:2016ijv}, but this is a topic we will address in future research.

\begin{acknowledgements}
T. T. Y. thanks Shigeki Matsumoto for the discussion on the possible suppression of the electron-positron coupling of the feeton.
This work is supported by 
the National Natural Science
Foundation of China (12175134, 12375101, 12090060, 12090064, and 12247141),
JSPS Grant-in-Aid for Scientific Research
Grants No.\,19H05810, 
the SJTU Double First Class start-up fund No.\,WF220442604,
and World Premier International Research Center
Initiative (WPI Initiative), MEXT, Japan.
\end{acknowledgements}

\providecommand{\href}[2]{#2}\begingroup\raggedright\endgroup

\vspace{15mm}
\end{document}